\documentclass[lettersize,journal]{IEEEtran}

\usepackage{bibentry}
\usepackage{amsthm}
\usepackage{amsmath,amssymb}
\usepackage{multirow}
\usepackage{makecell}
\usepackage{bm}
\usepackage{xspace} 
\usepackage{subfigure}
\usepackage{graphicx}
\usepackage{algorithm}
\usepackage{algorithmic}

\usepackage{amsmath}
\usepackage{amsthm}
\usepackage{booktabs}
\usepackage{algorithm}
\usepackage{algorithmic}
\usepackage{amsmath,amssymb}
\usepackage{booktabs}
\usepackage{multirow}
\usepackage{makecell}
\usepackage{bm}
\usepackage{xspace} 
\usepackage{subfigure}
\newcommand{\methodFont}{\textsl}
\newcommand{\RAGIC}{\methodFont{RAGIC}\xspace}

\begin{document}
\title{RAGIC: Risk-Aware Generative Adversarial Model for Stock Interval Construction}

\author{Jingyi~Gu,~\IEEEmembership{Student~Member,~IEEE,} Wenlu~Du,~\IEEEmembership{Student Member,~IEEE,}
        and~Guiling~(Grace)~Wang,~\IEEEmembership{Fellow,~IEEE}\\% <-this % stops a space
        \IEEEauthorblockA{\textit{Department of Computer Science} \\
\textit{New Jersey Institute of Technology}\\
Newark, New Jersey, United States \\
\{jg95@njit.edu, wd48@njit.edu, gwang@njit.edu\}}
}



\maketitle

\begin{abstract}
Efforts to predict stock market outcomes have yielded limited success due to the inherently stochastic nature of the market, influenced by numerous unpredictable factors. Many existing prediction approaches focus on single-point predictions, lacking the depth needed for effective decision-making and often overlooking market risk. To bridge this gap, we propose a novel model, \emph{RAGIC}, which introduces sequence generation for stock \emph{interval} prediction to quantify uncertainty more effectively. Our approach leverages a Generative Adversarial Network (GAN) to produce future price sequences infused with randomness inherent in financial markets.
\emph{RAGIC}'s generator includes a risk module, capturing the risk perception of informed investors, and a temporal module, accounting for historical price trends and seasonality. This multi-faceted generator informs the creation of \emph{risk-sensitive intervals} through statistical inference, incorporating \emph{horizon-wise} insights. The interval's width is carefully adjusted to reflect market volatility. Importantly, our approach relies solely on publicly available data and incurs only low computational overhead.
\emph{RAGIC}'s evaluation across globally recognized broad-based indices demonstrates its balanced performance, offering both accuracy and informativeness. Achieving a consistent 95\% coverage, \emph{RAGIC} maintains a narrow interval width. This promising outcome suggests that our approach effectively addresses the challenges of stock market prediction while incorporating vital risk considerations.
\end{abstract}


\section{Introduction}
\IEEEPARstart{T}{he} stock market is an essential component of financial systems, and stock prices reflect the dynamics of economic and financial activities. 
Predicting the future trend of individual stocks or overall market indices is always important to investors and other market players \cite{bollen2011twitter}, which requires significant efforts but lacks continuous success. 
Although some advanced methods based on machine learning and neural networks have recently been proposed to forecast stock price, return, or movement \cite{jiang2021applications}, they still face a main challenge: stock prices, especially broad-based indices (such as Dow30 and S\&P 500), carry inherent stochasticity caused by crowd emotional factors (such as fear and greed), economic factors (such as fiscal and financial policy and the economy of the US) and many other unknown factors \cite{malkiel1999random}.  

\begin{figure}
\centering \includegraphics[width=.5\textwidth]{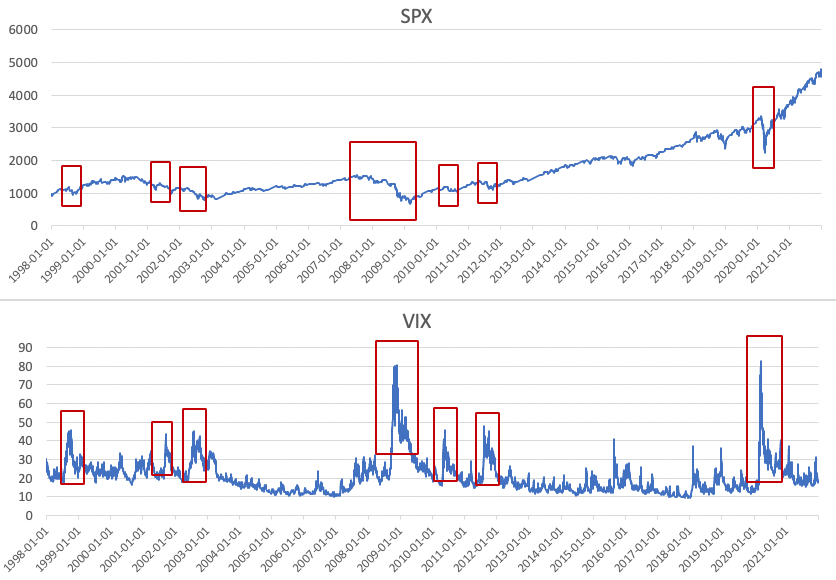}
  \caption{Illustration of the relationship between SPX and VIX in 1997-2021. The red areas in SPX show the price drop and the corresponding red areas in VIX show the peaks.}
  \label{fig:SPX_VIX}
\end{figure}

Despite big efforts to learn from various data, 
such as news, macroeconomic data, company fundamentals, sector data, and historical stock prices.
Existing works on stock market prediction suffer from two major limitations: 
(1) They mostly focus on \emph{point} prediction, which returns one single scalar for each time step.
However, point prediction is not informative enough to reflect market conditions driven by complex factors. 
Risk assessment and decision-making require quantification of uncertainty \cite{gollier2018economics} 
which is usually provided by \emph{interval} prediction \cite{stankeviciute2021conformal}, i.e., a price interval with lower and upper bound for each future time step. 
%Interval prediction in the financial market and even beyond is very under-explored, and it is the focus of this paper. Although diverse interval prediction methods have been proposed for time series, few works can be employed in the stock market directly due to the prohibitive computational cost, weak generalization ability, or strong assumption of independence. 
Interval prediction in the financial market, and even in other domains, remains significantly under-explored, making it the central focus of this paper. While several methods for interval prediction in time series have been proposed, they can not be directly applied to the stock market due to
their reliance on strong assumptions of independence, limited generalization capabilities, 
or high computational requirements. 
Moreover, widening the prediction interval increases coverage accuracy but reduces informativeness. 
It is challenging to balance accuracy and informativeness, and thus a proper setting needs to be carefully designed for the stock market ~\cite{yaniv1995graininess,lawrence2006judgmental}. 
(2) Most previous works largely neglect market risk. Accurate detection and early warning of risk enables hedging against black swan events, helps maintain market liquidity facing extreme circumstances, and stabilizes the financial market. The volatility index, derived from the prices of index options with near-term expiration dates, is often seen as a way to gauge market sentiment and evaluate the fluctuation risk. 
%An example of such evidence is shown in Figure \ref{fig:SPX_VIX}. The CBOE Volatility Index (VIX) is a real-time index representing the market's expectations for the relative strength of near-term price changes of the S\&P 500 index (SPX). The dramatic price drop of SPX in the financial crisis of 2009 and the COVID pandemic period corresponds to the peak of VIX in the same period. Therefore, the volatility index is essential for detecting the risk of fluctuation in the financial market.
One illustrative piece of evidence is presented in Figure \ref{fig:SPX_VIX}. The CBOE Volatility Index (VIX) is a real-time indicator reflecting the market's outlook on the expected magnitude of near-term price fluctuations in the S\&P 500 index (SPX). Notably, the substantial decline in SPX during the financial crisis of 2009 and the COVID pandemic aligns with the peak values observed in the VIX during the same periods. This correlation highlights the crucial role of the volatility index in detecting potential risks associated with market fluctuations.

To address the two major issues above, we propose \RAGIC, a Risk-Aware Generative model for Interval Construction. 
Specifically, \RAGIC has two phases: sequence generation and interval construction.
In the first phase, we employ a Generative Adversarial Network (GAN) to learn the historical stock features and simulate the future price sequences.
The generator incorporates a risk module, which captures risk using a \emph{risk attention score} derived from volatility index \footnote{Please note that the volatility index differs from volatility. Volatility is normally calculated as the square root of variance. But the stock volatility index is derived from multiple option prices of the underlying stock index. For example, VIX is the volatility index of SPX, and VXD is the volatility index of DOW.}, and a temporal module, which captures the multi-scale trend expressed by historical prices.
The well-trained generator can produce an adequate set of future price sequences 
with artificial randomness learned from the financial market. 
In the second phase, a \emph{horizon-wise strategy} is designed to gather simulated sequences with different prediction horizons, and statistical inference is utilized to construct a \emph{risk-sensitive interval} to reflect uncertainty, where the interval width is adaptively determined by the volatility index. 
In practice, \emph{RAGIC} is light-weighted to implement for its low computational cost and data availability. 
The GAN model is trained only once and can be repeatedly deployed to produce multiple sequences. Its reliance on public data only further enhances its accessibility.
Experiment results on multiple broad-based indices validate that our model achieves a balanced accuracy-informativeness manner, consistently providing 95\% coverage while maintaining a low width for intervals.

The major contributions are summarized as follows: 
\begin{itemize}
    \item To the best of our knowledge, we make the first attempt to employ a sequence generative model for interval prediction in stock markets, which quantifies uncertainty and provides much richer information regarding future market trends and risks. 
    \item We incorporate the volatility index to capture the risk perceived by smart money and designed a \emph{risk-sensitive interval} which adjusts the width of the predicted interval automatically. A wider interval indicates higher market volatility and a potentially bigger need for hedging. 
    \item Extensive experiments on multiple stock indices worldwide illustrate that \RAGIC achieves consistently over 95\% coverage with reasonable interval width in a balanced manner, significantly outperforming multiple popular benchmarks on interval prediction.

\end{itemize}
\section{Related Works}

\subsection{Stock Market Prediction}
Prediction task for stock market mainly has two categories: price or return prediction, which predicts a numerical future 
price or return, and movement prediction, which predicts a binary market up or down. 
Traditional methods include technical indicators and linear statistical models. 
Recently, many deep neural networks have been employed \cite{wang2021coupling,lin2021learning,li2019dp,lee2017predict,gu2023stock,gu2023deep,ye2023prediction}
. To name a few, DARNN \cite{qin2017dual} proposes a dual-stage attention-based RNN model to predict the stock trend; AdvLSTM \cite{feng2018enhancing} employs the concept of adversarial training on LSTM \cite{hochreiter1997long} to predict stock market price; HMG-TF \cite{ding2020hierarchical} designs an enhanced version of Transformer; 
FactorVAE \cite{duan2022factorvae} integrates Variational Autoencoder (VAE) and dynamic factor model to predict returns. 
There have been few works on predicting future stock price range. 
Among them, linear regression with ordinary least square \cite{hu2007application} is adopted to forecast interval for price. 
Bootstrap method \cite{de2020construction} is applied to build a confidence interval for mean price. 
Even though some works attempted to learn from noisy data, they still have neglected risk, even when facing market uncertainty. 
To this end, our work aims to incorporate market volatility and quantify the uncertainty of stock by interval prediction.

\subsection{Interval Prediction on Time Series}

Uncertainty quantification is often formulated into predicting intervals. 
Previous work related to neural networks can be categorized into four types for dynamic time series. 
(1) Bayesian methods  \cite{mackay1992practical}  apply prior and posterior probability distribution on the parameters of neural networks to quantify the uncertainty \cite{neal2012bayesian}. 
The approximated Bayesian inference with dropout training \cite{gal2016dropout} is used in practice.
(2) Quantile methods extend quantile regression \cite{koenker2001quantile} to neural networks. 
They model the cumulative distribution of the target and predict interval boundary directly \cite{wen2017multi, khosravi2010lower}, 
while being limited by generalization ability to new tasks and datasets.
(3) Ensembles methods combine a set of models trained with bootstrapped samples\cite{shrestha2009novel}, different configurations of parameters or model structures \cite{fort2019deep}. 
However, such methods have a prohibitive computational cost on time and space.
(4) Conformal prediction methods \cite{shafer2008tutorial} use a nonconformity measure for calibration and derive the prediction interval from a set of predictions with residual errors \cite{stankeviciute2021conformal}. 
They assume the exchangeability of data. 
However, real-world time series hardly satisfy such an assumption due to temporal dependency.
To quantify the uncertainty in the stock market, we construct the interval by combining multiple sequences simulated 
from a generative model. Our work is efficient on time and space since the model is trained only once, 
and it can well capture temporal relations in stock price series.

\subsection{Generative Adversarial Nets on Time Series Data}
Emerging works use GAN on time series data in video generation, text generation, biology, finance, music, and many other applications \cite{zhang2016generating, saito2017temporal, dong2018musegan, yu2017seqgan,yoon2019time}. 
These GANs aim to generate a synthetic sequence that retains the characteristics of the original sequence.
For example, TimeGAN\cite{yoon2019time} is proposed to generate the realistic time series, which combines an unsupervised paradigm with supervised autoregressive models.  They are mostly employed for
tasks such as enriching limited real-world datasets \cite{wiese2020quant}, data imputation \cite{xu2020scigans}, and machine translation \cite{wu2018adversarial}. 
% SeqGAN \cite{yu2017seqgan} extends GANs with reinforcement learning to generate sequences of discrete tokens.  
However, stock market prediction is to discover future patterns instead of historical time series. 
Thus, most existing efforts of time series GANs are distinct from our objective and cannot be used directly.
Some recent approaches implement GANs to predict stock trends \cite{zhang2019stock,zhang2016generating}. 
Nevertheless, their simple structure fails to capture the inherent temporal patterns of stock trends, 
and vanilla GAN with one-step prediction hardly achieves superior performance. 
In this paper, we fully explore the potential of GAN to generate the synthetic sequence and predict stock price multi-step forward.

\section{Problem Formulation}

The paper aims to provide stock interval prediction. 
The problem is formulated into two phases: sequence generation and interval construction.\footnote{For notations, we use bold letters for vectors or matrices and plain letters for scalars, if not specified otherwise.}
(1) In the first phase, we aim to develop a model to generate sequences of future stock prices. Let $\bm{x}=(\bm{x}_1,...\bm{x}_W)\in \mathbb{R}^{W\times K}$ denote the historical stock characteristics with $W$ time steps and $K$ features. The stock features consist of price-based features $\bm{p}$ and volatility-based features $\bm{v}$. Let $\bm{z} \in \mathbb{R}^W$ denote a random noise sample which is assumed to be drawn from a uniform distribution. Given $\bm{x}$ and $\bm{z}$, our task is to learn a generative adversarial model $f$ with parameter $\Theta$ to predict the sequence of close price $\bm{y}=(y_{1},...,y_{H})\in \mathbb{R}^H$ in the future $H$ steps.
(2) In the second phase, the well-trained generative model can produce a set of future price sequences $\hat{\bm{Y}}=(\hat{\bm{y}}_{1},...,\hat{\bm{y}}_{N}) \in \mathbb{R}^{N \times H}$ by repeatedly feeding stock features and different noise for $N$ iterations: 
\begin{equation}
    \hat{\bm{y}}_{n} = f(\bm{x},\bm{z}_n;\Theta), n\in {1,...,N}
\end{equation}
where $\bm{z}_n$ is drawn from distribution randomly and independently for $n^{th}$ iteration.
Given the generated sequences $\hat{\bm{Y}}$, the goal is to build a prediction interval in the form of $[Y_t^L, Y_t^U]$ such that it accurately covers future actual price $y_t$, where $Y_t^L$ and $Y_t^U$ are lower and upper bound respectively. 
A prediction interval with a narrow width $Y_t^U-Y_t^L$ 
and high accuracy in covering the actual price is desired.

\section{Methodologies}
% The proposed framework includes two phases.
% The first phase is sequence generation, where a GAN model is trained to learn patterns on stock features and generate the predicted sequences of future price, as shown in Figure~\ref{fig:overall_structure}. During the training process,
% a generator tailored for the stock market is designed with a risk module and a temporal module to predict the future sequence, and a critic approximates the Wasserstein distance between the real and predicted sequences, which is utilized for optimization. 
% With the well-trained GAN, the generator can produce a set of future price sequences with uncertainty by feeding different noises and stock features. The sequences can be seen as a simulation of future prices with all possible uncertainty from the financial market. 
% The second phase is the novel interval construction, as shown in Figure \ref{fig:model_interval}. The \emph{risk-sensitive interval} is built using statistical inference on the simulated sequences from the first phase with \emph{horizon-wise} information. Meanwhile, the interval width is adapted to market risk. Compared with other state-of-the-art methods, the sequences can be obtained in minimal time without a second training. 

\subsection{Sequence Generation}

The general form of GAN consists of two networks. The generator $\mathcal{G}$ learns real sequence $\bm{y}_t$. Specifically, a random noise sample $z$, together with stock features $\bm{x}_t$ as the condition to control the mode of generated sequence, are fed into generator $\mathcal{G}$ to build a synthetic sequence $\hat{\bm{y}}_t$. 
The critic $\mathcal{C}$, also known as the discriminator, is an appropriate measure of distance. 
In this paper, we choose Wasserstein distance from WGAN \cite{arjovsky2017wasserstein} to measure the difference between the real and generated data from the perspective of optimal transportation problem \cite{arjovsky2017wasserstein}.
Intuitively, it indicates the cost or mass caused by transporting $\bm{y}$ to $\hat{\bm{y}}$ \cite{rubner2000earth}. 
In practice, Kantorovich-Rubinstein duality is used to reconstruct the WGAN value function in critic and solve the optimization problem:
\begin{equation}
\min\limits_{\mathcal{G}} \max\limits_{\mathcal{C} \in \mathcal{F}} \mathbb{E}_{\bm{y} \sim P_r} [\mathcal{C}(\bm{y})] - \mathbb{E}_{\hat{\bm{y}} \sim P_g} [\mathcal{C}(\hat{\bm{y}})]
\end{equation}
where $P_g$ and $P_r$ are implicitly defined by $\hat{\bm{y}} $ and $\bm{y}$ respectively, $\mathcal{F}$ is the set of 1-Lipschitz functions. The generator $\mathcal{G}$ and critic $\mathcal{C}$ are trained simultaneously. The generated sequence from the generator is enforced to be as close as possible to the real sequence. The critic aims to find the optimal value function to better approximate the Wasserstein Distance. In the inference phase, generator $\mathcal{G}$ produces a fake sequence from the random noise $\bm{z}$ and condition $\bm{x}$.

\begin{figure*}
  \centering
  \includegraphics[width=.99\textwidth]{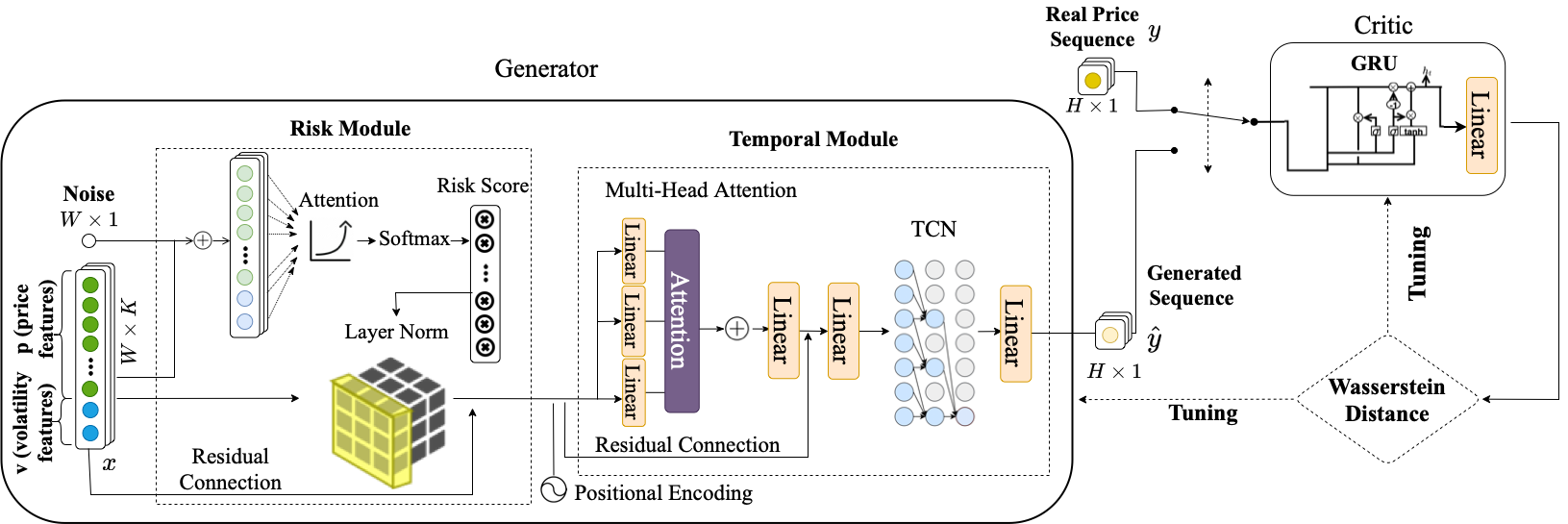}
  \caption{The overall architecture of GAN in \RAGIC during the training process.}
  \label{fig:overall_structure}
\end{figure*}

\subsubsection{Generator}
% Our generator consists of two parts: a risk module and a temporal module. 
% In addition to stock features, a random noise sample is added as an input feature to simulate the stochasticity nature of stock price. 
% For notational convenience, we absorb it into $K$, the dimension of stock features $\bm{x}$. 
% The historical stock features are fed into the generator. After flowing through the risk module and the temporal module, the generator outputs predicted future price sequences.

\paragraph{Risk Module}
Market volatility implies investment risk.
%, representing investors' expectation of near-term market fluctuation. 
Higher volatility index indicates a declining market with a higher probability, 
while a low one may imply a rising market. 
To capture the risk perceived by smart money, we design a risk score to enhance the features.
% $\bm{x}$ consists of price-based features $\bm{p}$ and volatility-based features $\bm{v}$. 
For each feature $\bm{x}_k\in\mathbb{R}^W, k\in(1,...K)$, a \emph{risk attention score} $\bm{\beta}_k\in \mathbb{R}^W$ is computed:

\begin{equation}
    \bm{\beta}_k=
    \begin{cases}
    \exp{(\bm{\lambda}_k \max(0,\bm{x}_k - \bm{\delta}_k))} & \bm{x}_k\in\bm{v} \\
    1 & \bm{x}_k\notin\bm{v} \\
    \end{cases}
\end{equation}
For volatility-based features $\bm{v}$, if its value exceeds the risk threshold $\bm{\delta}_k\in \mathbb{R}^W$, then the current market can be seen as with higher risk and the corresponding \emph{risk attention score} $\bm{\beta_k}$ will grow exponentially. The coefficient $\bm{\lambda}_k\in \mathbb{R}^W$ controls the growth rate. Otherwise, $\bm{\beta}_k$ is set to 1, the same as those of priced-based features $\bm{p}$. 
Then we normalize the \emph{risk attention score} by $\bm{r}_k=softmax(\bm{\beta}_k)$ and assign it to corresponding feature $\bm{x}_k$. Additionally, We adopt residual connections and layer normalization \cite{xiong2020layer} for the purpose of faster training: $\tilde{\bm{x}}_k=\bm{x}_k+LayerNorm(\bm{r}_k \odot \bm{x}_k)$, where $\odot$ is element-wise multiplication, $\tilde{\bm{x}_k}\in\mathbb{R}^W$  is the risk-enhanced features.
The output of risk module is $\tilde{\bm{x}}=( \tilde{\bm{x}}_1,..., \tilde{\bm{x}}_k)\in\mathbb{R}^{W \times K}$,  which is obtained by concatenating all enhanced features together. 

\paragraph{Temporal Module}
To adaptively capture the temporal dependency and seasonality of the stock market and generate synthetic sequences, 
we design a temporal module. Firstly, we add positional encoding \cite{vaswani2017attention} to update the risk-enhanced features $\tilde{\bm{x}} = \tilde{\bm{x}} + Positional Encoding(\tilde{\bm{x}})$. We then employ multi-head attention to learn the interrelationship of different historical time steps globally. First, the risk-enhanced features $\tilde{\bm{x}}$ is linearly mapped into query $\bm{q}_d= \tilde{\bm{x}}\bm{W}_d^q\in\mathbb{R}^{W \times d_k}$, key $\bm{k}_d= \tilde{\bm{x}}\bm{W}_d^k\in\mathbb{R}^{W \times d_k}$ and value $\bm{v}_d= \tilde{\bm{x}}\bm{W}_d^v\in\mathbb{R}^{W \times d_k}$ in each head, where $d\in(1,..,D)$ is the head index, $d_k$ is hidden dimension,
$\bm{W}_d^q$, $\bm{W}_d^k$, and $\bm{W}_d^v$ are weight parameters. Attention score  $\bm{\phi}_h\in\mathbb{R}^{W \times W}$ and
the output $\bm{o}_d\in\mathbb{R}^{W\times d_k}$ from $d^{th}$ head can be calculated:
\begin{equation}
    \bm{o}_d = softmax(\bm{\phi}_d)\bm{v}_d, \quad
    \bm{\phi}_d = \frac{\bm{q}_d \bm{k}_d^\top}{\sqrt{d_k}}
\end{equation}
Each element of the attention score 
$\bm{\phi}_{d, (i,j)}$ represents the interrelationship of features between time step $w_i$ and $w_j$. We concatenate the outputs from all the heads together, then employ a linear mapping layer and a residual connection to generate final output $\bm{s}= \tilde{\bm{x}} + [\bm{o}_1,...,\bm{o}_d] \bm{W}\in\mathbb{R}^{W \times K}$.
%Grace: pls double check the last sentence. 

Moreover, stock data exhibits seasonal patterns across various timeframes. For instance, analyzing weekly data highlights lower S\&P 500 returns on Mondays compared to Wednesdays \cite{french1980stock}. Monthly trends further reveal heightened volatility and volume due to options contracts expiring on the third Friday \cite{chiang2014stock}. Additionally, quarterly earnings seasons and U.S. Federal Reserve Federal Open Market Committee (FOMC) meetings contribute to increased market volatility \cite{mian2012investor}.
To capture such trends, we adopt dilated convolution layer in Temporal Convolution Network (TCN) \cite{bai2018empirical} which employs a 1D fully convolutional network and causal convolutions.
Specifically, the dilated convolution representation $\tilde{\bm{s}}_w^l$ at the $w^{th}$ time step and $l^{th}$ layer is calculated from the input $\bm{s}$ as follows:

\begin{gather}
    \tilde{\bm{s}}_w^{(l)} = \bm{f}  \circledast (\tilde{\bm{s}}_w^{(l-1)},...,\tilde{\bm{s}}_{w-\eta p}^{(l-1)} )\\
    \tilde{\bm{s}}_w^{(1)} = \bm{f}  \circledast (\bm{s}_w,...,\bm{s}_{w-\eta p})
\end{gather}
where $l\in(1,...,L)$ is the layer index, $\bm{f}\in\mathbb{R}^p$ is the kernel with size $p$, $\circledast$ is the convolution operation between kernel and sequence element, $\eta=2^{l-1}$ is the dilation rate. When $\eta=1$, it is equivalent to a regular convolution layer. As the dilation rate grows in the following layer exponentially, the hidden output at the top layers receives a wider range of historical information. This scheme is able to learn longer periodical patterns without information loss. After obtaining the hidden output from the top layer, we adopt a fully connected layer to generate fake sequence $\hat{\bm{y}}\in\mathbb{R}^{H}$. 

\subsubsection{Critic}

The critic is designed to approximate the Wasserstein distance between real and generated sequences, which is used for optimization during the training. 
Let $\bm{y}'$ denote either a real or fake sequence as the input to the critic. We use Gated Recurrent Units (GRU), which is a simple yet effective structure, to temporally capture the dependency and a fully connected layer to pass the last hidden state, and obtain the approximated distance value $c=\mathcal{C}(\bm{y}';\Theta_c)= FC(GRU(\bm{y}'))$. $\Theta_c$ represents the set of parameters in $\mathcal{C}$.

\subsubsection{Optimization}

Solely relying on adversarial feedback is insufficient for the generator to capture the conditional distribution of data. To improve the similarity between two distributions more efficiently, we introduce an additional loss on the generator to discipline the learning. Specifically, the supervised loss $\mathcal{L}_S$ is to directly describe the $L1$ distance between $\bm{y}$ and $\hat{\bm{y}}$. By integrating the supervised loss with adversarial feedback, the objective functions for the generator and critic networks are presented:

\begin{gather}
\mathcal{L}_\mathcal{G} = -\mathbb{E}_{\hat{\bm{y}} \sim P_g} [\mathcal{C}(\hat{\bm{y}})] + \gamma \mathcal{L}_S \\
\mathcal{L}_S = \mathbb{E}_{\hat{\bm{y}} \sim P_g, \bm{y} \sim P_r} [\parallel \bm{y}-\hat{\bm{y}}\parallel]
\\
\mathcal{L}_\mathcal{C} = -\mathbb{E}_{\bm{y} \sim P_r} [\mathcal{C}(\bm{y})] + \mathbb{E}_{\hat{\bm{y}} \sim P_g} [\mathcal{C}(\hat{\bm{y}})]
\end{gather}
where  $\gamma \geq 0$ is the penalty parameter on supervised loss.

To guarantee the Lipschitz constraint on Critic, it's necessary to force the weights of the critic to lie in a compact space. In the training process, we clip the weights to a fixed interval $[-\xi, \xi]$ after each gradient update.

\begin{figure*}[!htp]
    \centering 
    \subfigure{      \includegraphics[width=.6\textwidth]{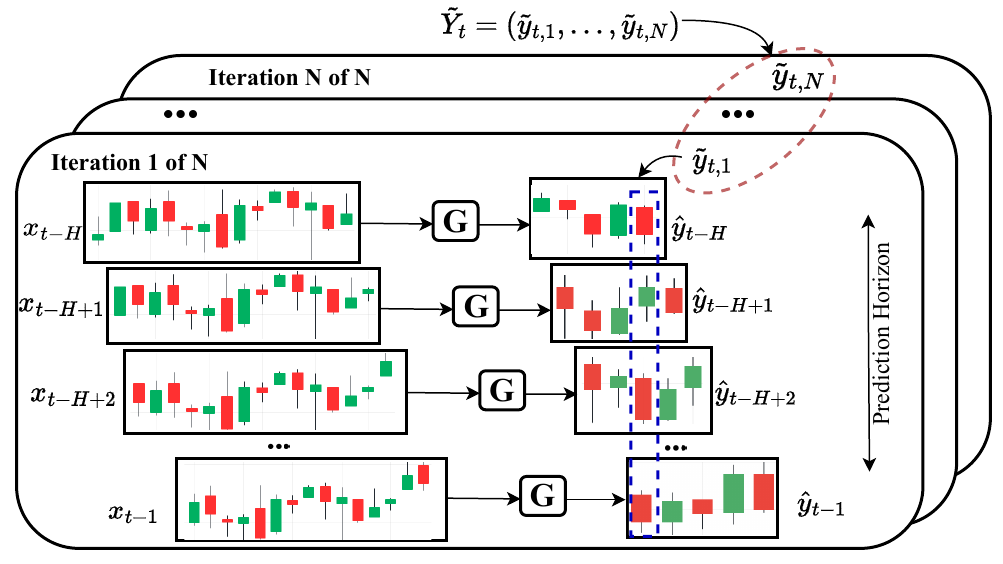}}\subfigure{\includegraphics[width=.4\textwidth]{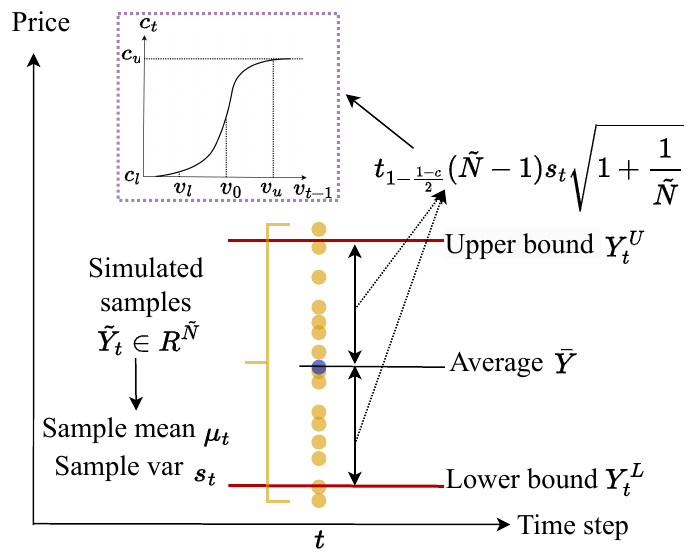}}
    \caption{Sequence simulation (left) and \emph{risk-sensitive interval} construction (right).}
    \label{fig:model_interval}
\end{figure*}

\subsection{Interval Construction}

The GAN model is trained just once, then it can be deployed repeatedly to produce multiple sequences that can be utilized to construct intervals.
The interval construction consists of  \emph{horizon-wise} sequence simulation and \emph{risk-sensitive interval} construction.

\subsubsection{Sequence Simulation} 
The generated sequence $\hat{\bm{y}}$ from generator covers $H$ steps ahead, i.e., 
we have $H$ future prices predicted at different times. 
To fully utilize this horizon information, we propose a \emph{horizon-wise strategy}. 
Specifically, to predict the target at time step $t$, our generator is executed for $H$ rounds. 
For each round $j\in(1,...,H)$, the noise vector $\bm{z}$ and historical features $\bm{x}_{t-j}=(\bm{x}_{t-j-W+1},...,\bm{x}_{t-j})$
are fed into generator to generate sequence $\hat{\bm{y}}_{t-j}=(\hat{y}_{t-j+1},..., \hat{y}_{t-j+H})\in\mathbb{R}^H$. We denote $\hat{y}^j_{t}=\hat{y}_{t}$ as prediction at $j^{th}$ position. 
An example of this prediction vector is marked by the dotted blue line in Figure \ref{fig:model_interval}. 
To predict the interval for a particular day $t$, we can obtain a prediction vector $\tilde{\bm{y}}_t=(\hat{y}^1_t ,..., \hat{y}^H_t)\in\mathbb{R}^H$ from $H$ sequences. 

We repeat the above operation for $N$ times to capture the market randomness by feeding different noises and stock features into the model. 
Thus, $N$ different prediction vectors for time step $t$ are generated. In total, for time $t$, we have $\tilde{N}=N\times H$ different predictions. 
All these predictions can be seen as a simulation of future trends with diverse perspectives in different horizons. They thus can well capture and express the unpredictable uncertainty of the market.  
The concatenation of prediction vectors is denoted as $\tilde{\bm{Y}_t}=(\tilde{\bm{y}}_{t,1},...,\tilde{\bm{y}}_{t,N})\in\mathbb{R}^{\tilde{N}}$.

\subsubsection{Risk-Sensitive Interval}

We aim to build a \emph{risk-sensitive prediction interval} to accurately capture the possible price range with a dynamic width determined by the perceived market risk and fluctuation.

First, we utilize statistical inference to build the prediction interval. The stock price at an individual time step is assumed to be a random variable $Y_t$ which follows Gaussian distribution with the unknown population mean $\mu_t$ and variance $\sigma^2_t$ at this time step, denoted as $Y_t\sim \mathcal{N}(\mu_t, \sigma^2_t)$. 
Given the simulated set $\tilde{\bm{Y}_t}$ for time step $t$, the sample mean $\Bar{Y}_t$ and sample variance $s_t^2$ are calculated as:
\begin{equation}
    \Bar{Y}_t = \frac{1}{\tilde{N}}\sum_{i=1}^{\tilde{N}}\tilde{y}_{t,i}, \quad
    s_t^2 = \frac{1}{\tilde{N}-1}\sum_{i=1}^{\tilde{N}}(\tilde{y}_{t,i}-\Bar{Y}_t)^2
\end{equation}
where $\tilde{y}_{t,i}$ is $i$-th sample in the set, $\tilde{N}-1$ of $s_t^2$ is for unbias correction. 

%Grace: please double check this part: Y_N+1 is a point following the t distribution 
% Jingyi: Y_N+1 follows normal distribution.  we can infer the transformation of it follows t distribution. $\frac{Y_{t,\tilde{N}+1}-\Bar{Y}_t}{s_t\sqrt{1+\frac{1}{\tilde{N}}}} \sim t_{(\tilde{N}-1)}$. 
%got it. u can just double check this paragraph 
% for this explanation on distribution, I may prefer to use previous version. 
Thus, we can infer that $\frac{Y_{t}-\Bar{Y}_t}{s_t\sqrt{1+\frac{1}{\tilde{N}}}} \sim t_{(\tilde{N}-1)}$ empirically. \footnote{Note that $t$ in $t_{(\tilde{N}-1)}$ represents the t-distribution; $t$ as subscript such as $s_t$ represents the time step.}
% Thus, $Y_{\tilde{N}+1}$ follows distribution $\frac{Y_{t,\tilde{N}+1}-\Bar{Y}_t}{s_t\sqrt{1+\frac{1}{\tilde{N}}}} \sim t_{(\tilde{N}-1)}$.
% \begin{equation}
%     \frac{Y_{t,\tilde{N}+1}-\Bar{Y}_t}{\sigma_t\sqrt{1+\frac{1}{\tilde{N}}}} \sim \mathcal{N}(0,1)   ,\quad
%     \frac{s_t^2}{\sigma_t^2}\sim\chi^2_{(\tilde{N}-1)}
% \end{equation}
%Then we could calculate a prediction interval $[Y_t^L$, $Y_t^U]$ with $Y_t^L$ and $Y_t^U$ as lower and upper bound respectively, as shown in Figure \ref{fig:model_interval} (left):
Accordingly, the prediction interval $[Y_t^L$, $Y_t^U]$ with $Y_t^L$ and $Y_t^U$ as lower and upper bound, can be calculated:
\begin{gather}
    Y_t^L = \Bar{Y}_t - t_{1-\frac{(1-c)}{2}}(\tilde{N}-1)s_t\sqrt{1+\frac{1}{\tilde{N}}}\\
    Y_t^U = \Bar{Y}_t + t_{1-\frac{(1-c)}{2}}(\tilde{N}-1)s_t\sqrt{1+\frac{1}{\tilde{N}}}
\end{gather}
The probability of interval containing the true value equals confidence $c\in[0,1]$:
\begin{equation}
    P(Y_t^L\leq Y_{t}\leq Y_t^U)=c
\end{equation}
where $t_{1-\frac{(1-c)}{2}}(\tilde{N}-1)$ is the $(1-\frac{(1-c)}{2})^{th}$ percentile of t-distribution with $\tilde{N}-1$ degrees of freedom. The prediction interval gets wider with increasing $c$. The most commonly used values for confidence $c$ are 90\%, 95\%, and 99.9\%. 
The procedure is illustrated in Figure \ref{fig:model_interval} (left). More details of the statistical inference are in Supplementary.

% table for interval performance

Instead of using a fixed $c$, we propose to adapt $c$ to the market risk.
Intuitively, when the market is in a riskier condition, 
the interval should be wider to capture the market's higher uncertainty; 
otherwise, a narrow interval is adequate. 
Hence, we design a \emph{risk-sensitive interval} construction, market risk determines $c$ and
and further controls the percentile of t-distribution as well as the width of the predicted interval accordingly.
$v_l, v_u$ are defined as the lower and upper thresholds for the volatility features. $c_t$ is limited in the range $[c_l,c_u]$. It shows a sigmoid curve $F$ between $c_t$ and volatility feature $v_{t-1}$, as shown in Figure \ref{fig:model_interval} (right):
\begin{gather}
    c_t = F(v_{t-1})= c_l+\frac{c_u - c_l}{1+\exp{(-k(v_{t-1}-v_0))}} \\
    k =\ln{(\frac{c_u-c_l}{\Delta \epsilon}-1})\frac{2}{v_u-v_l}
    ,\quad v_0=\frac{v_l+v_u}{2}
\end{gather}
where $k$ is the growth rate showing the steepness of the curve, and $v_0$ is the middle point of the curve, as well as middle point of lower and upper volatility thresholds. 
At first, $c_t$ stays close to $c_l$ infinitely but not equal to $c_l$ when $v_{t-1} \leq v_l$, with at most $\Delta\epsilon$ of difference; when $v_{t-1}$ increases from $v_l$ to $v_u$, $c_t$ converges quickly from $c_l$ to $c_u$; as $v_{t-1}$ approaches to and exceeds $v_u$, $c_t$ stay close to $c_u$ infinitely. In this paper, $c_l$ and $c_u$ are set to 90\% and 99.9\%.
% Since $c$ controls percentile of t-distribution, the risk-alpha function automatically changes interval width for the time step $t$ based on the market risk at the previous time step.
Hence, a wider predicted interval indicates higher market volatility and potentially bigger needs for hedging.

\section{Training and Inference Algorithms}

\subsection{Training Process}

Algorithm \ref{alg:training} presents the details of the learning process for the generator $\mathcal{G}$ and critic $\mathcal{C}$ in \RAGIC during the sequence generation phase. Let $\mathcal{D}_{train}$ denote the training data containing input features $\bm{x}$ and real sequence $\bm{y}$, and let $\mathcal{D}_{batch}$ denote a subset of $\mathcal{D}_{train}$ for the training in each batch. 
% All the equations mentioned in the Algorithm \ref{alg:training} are in the main paper.

% algo for training process

\begin{algorithm}
\caption{Training Process during Sequence Generation}
\textbf{Input}: Training data $\mathcal{D}_{train}$, generator $\mathcal{G}$, critic $\mathcal{C}$ \\
\textbf{Output}: Well-trained models $\mathcal{G}$ and $\mathcal{C}$ with optimal parameters $\Theta_\mathcal{G}^*$, $\Theta_\mathcal{C}^*$ \\
\textbf{Parameter}: model parameters: $\Theta_\mathcal{G}$, $\Theta_\mathcal{C}$, penalty $\gamma$, clipping threshold $\xi$, the number of epochs $n_{epochs}$, the number of iterations of critic per generator iteration $n_{critic}$ \\

\begin{algorithmic}[1] %[1] enables line numbers
\STATE Initialize $\mathcal{G}$, $\mathcal{C}$, $i\gets 0$.
\FOR{$epoch \in \{1,\dots,n_{epochs}\}$}
    \FOR{$\mathcal{D}_{batch} \gets \mathcal{D}_{train}$}
        \STATE $\bm{x}, \bm{y} \gets \mathcal{D}_{batch}$.  %\COMMENT{Train critic}
        \STATE Sample a batch of random noise $\bm{z}$ from uniform distribution.
        \STATE $\hat{\bm{y}}=\mathcal{G}(\bm{x},\bm{z};\Theta_\mathcal{C})$.
        \STATE Compute  $\mathcal{C}(\hat{\bm{y}};\Theta_\mathcal{G})$ and $\mathcal{C}(\bm{y};\Theta_\mathcal{C})$.
        \STATE Compute $\mathcal{L}_{\mathcal{C}}$ by Eq.9.
        \STATE Update $\Theta_\mathcal{C}$ in $\mathcal{C}$ by minimizing $\mathcal{L}_\mathcal{C}$.
        \STATE Clip weights of $\Theta_\mathcal{C}$ into $[-\xi, \xi]$.
        \STATE $i \gets i+1$.
        
        % Train generator every $n_critic$ iterations
        \IF{$i \% n_{critic} == 0$}
            \STATE $\hat{\bm{y}}=\mathcal{G}(\bm{x},\bm{z};\Theta_\mathcal{G})$.
            \STATE Compute  $\mathcal{C}(\hat{\bm{y}};\Theta_\mathcal{C})$.
            \STATE Compute $\mathcal{L}_{\mathcal{G}}$ by Eq.7.
            \STATE Update $\Theta_\mathcal{G}$ in $\mathcal{G}$ by minimizing $\mathcal{L}_\mathcal{G}$.
        \ENDIF
\ENDFOR
\ENDFOR
\end{algorithmic}
\label{alg:training}
\end{algorithm}

\subsection{Inference Phase}

Algorithm \ref{alg:inference} shows the details of the inference phase of \RAGIC.  Let $\mathcal{D}_{test}$ denote the test data containing a set of $H$ consecutive input features $(\bm{x}_{t-H},\dots,\bm{x}_{t-1})$ and real sequence $(\bm{y}_{t-H},\dots,\bm{y}_{t-1})$. 
% The equations mentioned are in the main paper.

\begin{algorithm}
\caption{Inference Phase for Interval Construction}
\textbf{Input}: Testing data $\mathcal{D}_{test}$, well-trained generator $\mathcal{G}$, \\
\textbf{Output}: Predicted interval $[Y_t^L, Y_t^U]$ at time step $t$\\
% and point $\hat{y}_t$ for price \\
\textbf{Parameter}: the number of sequences $N$ \\

\begin{algorithmic}[1] %[1] enables line numbers
\STATE \textbf{Initialize} $\tilde{\bm{Y}}_t=\{\}$, $v_{t-1}=0$, $\alpha=0$.
\FOR {$j \in \{H,\dots,1\}$}
    \STATE $\bm{x}_{t-j}, \bm{y}_{t-j} \gets \mathcal{D}_{test}$.
    \FOR {$n \in \{1,\dots,N\}$} 
    \STATE $\hat{\bm{y}}_{t-j,n} = \mathcal{G}(\bm{x}_{t-j}, \bm{z}_n;\Theta_\mathcal{G})$.
    \STATE $\tilde{\bm{Y}}_t \gets \tilde{\bm{Y}}_t \cup j$-th element in $\hat{\bm{y}}_{t-j,n}$.
    \ENDFOR
    \IF {$j==1$}
        \STATE $v_{t-1}\gets v_{t-1}$ in $\bm{x}_{t-1}$.
    \ENDIF
\ENDFOR
\STATE Compute $c_t$ by Eq.14.
\STATE Compute predicted interval bounds $Y_t^L$, $Y_t^U$ by Eq.11-12.
% \STATE Compute $\hat{y}_{t,n}$ by Eq.19.
% \STATE Compute predicted point $\hat{y}_t$ as average of $\{\hat{y}_{t,n}\}_{n=1}^{N}$.
\STATE \textbf{Return} $Y_t^L$, $Y_t^U$.
% , $\hat{y}_t$.
\end{algorithmic}
\label{alg:inference}
\end{algorithm}

\section{Model Evaluation}

\subsection{Experiment 1: Interval Prediction}
\subsubsection{Dataset}
We choose five broad-based stock indices worldwide, including three indices from the U.S. market, one index from the European market, and one index from Asian market: 
\begin{itemize}
    \item \textbf{Dow Jones Industrial Average (DJIA)} is one of the oldest equity indices, tracking 30 of the most highly capitalized and influential companies in major sectors of the U.S. market
    \item \textbf{S\&P 500 (SPX)} is one of the most commonly followed indices and is composed of the 500 largest publicly traded companies
    \item \textbf{NASDAQ 100} contains 101 of the largest non-financial company stocks from Nasdaq stock exchange; \footnote{Google has both GOOG, class C stock, and GOOGL, class A stock, listed in Nasdaq.}
    \item \textbf{Deutscher Aktienindex performance index (DAX)} measures the performance of 40 major German blue chip companies
    \item \textbf{The Nikkei index} consists of 225 companies from a wide range of industry sectors in Japan.
\end{itemize}

% All features in our model are derived from 
% The historical daily prices of stock indices (open, close, high, low) and their corresponding volatility indices are publicly available on Yahoo Finance \footnote{\url{https://finance.yahoo.com}}
The historical daily stock indices and their corresponding volatility indices are publicly available on Yahoo Finance \footnote{\url{https://finance.yahoo.com}} and Investing \footnote{\url{https://www.investing.com}}, and thus our model is easy to implement.

All indices are split into training, validation, and testing sets along with the temporal dimension. Even though the start date of the whole period varies in indices due to data availability, the validation period is consistently chosen as 2013/8-2015/12, and the testing period is consistently chosen as 2016/1-2021/12. 
Specifically, the testing period includes over 1400 trading days for all indices. It covers the COVID pandemic in 2020 and the following recovery when the market was highly volatile. We use features on the previous 30 days to generate sequences on the future 5 days, i.e., $W=30, H=5$. Table \ref{tab:time} shows the detailed statistics and training period of all indices.

\begin{table}[!htbp]
    \centering\fontsize{9}{9}\selectfont
    \caption{Statistics of Stock Indices}
    \begin{tabular}{cccc}
    \toprule   Index  & Volatility & \# days & Time Period of Training Set \\
    \midrule  
    DJI & VXD & 6051 & 1997/12-2013/7 \\
    SPX & VIX & 6051 & 1997/12-2013/7 \\
    Nasdaq & VXN & 5259 & 2001/1-2013/7 \\
    DAX & VDAX & 5237 & 2001/5-2013/7 \\
    Nikkei & JNIV & 4791 & 2002/6-2013/7\\
    \bottomrule 
    \end{tabular}
    
    \label{tab:time}
\end{table}

The input features of \RAGIC include price-based features and volatility-based features. The price-based features consist of historical raw prices (open, high, low, close) and eight deliberately chosen technical indicators that can be derived from raw prices, as shown in Table \ref{tab:feature}. For feature preprocessing:
\begin{itemize}
    \item Raw prices: converted to return calculated by $x_t / x_{t-1}-1$.
    \item Volatility: two features are derived: (1) volatility return  $v_t \ v_{t-1}-1$ is calculated the same as raw price features; (2) normalized volatility min-max scaled. The risk module uses both features as inputs, and interval construction is based on the raw volatility index.
\end{itemize}
 
In the phase of interval construction, the close return is converted back to the close price for sequence simulation, and the confidence is determined by the raw volatility index. 
More details on processing are in Supplementary.

\begin{table}[!htbp]
    \centering\fontsize{9}{9}\selectfont
    \caption{Input features.}
    \begin{tabular}{c|c}
    \hline
    Type & Features\\
    \hline
    \multirow{2}*{Price} & Open, Close, High, Low   \\
     & Volume \\
    \hline
    \multirow{6}*{Technical indicator} 
    & MACD DIFF  \\
    & Upper Bollinger band indicator  \\
    & Lower Bollinger band indicator  \\
    & EMA: 5-day   \\ 
    &SMA: 13-day, 21-day, 50-day \\
    \hline
    \multirow{2}*{Volatility} & Volatility index \\
    & Return of Volatility index \\ 
    \hline
    \end{tabular}
    \label{tab:feature}
\end{table}

\subsubsection{Baselines}

We select the following methods as our baselines for interval prediction:
\begin{itemize}
    
    \item \textbf{Bollinger Bands}  \cite{bollinger2002bollinger} is a momentum indicator in finance. 
    The upper and lower bands provide the oversold or overbought signals in the market.
    
    \item \textbf{BayesianNN} \cite{gal2016dropout} uses Monte Carlo dropout as Bayesian inference approximation.
    
    \item \textbf{MQRNN} \cite{wen2017multi} is a sequence-to-sequence framework for multi-horizon quantile forecasting.
    
    \item \textbf{CFRNN} \cite{stankeviciute2021conformal} is a neural network extending inductive conformal prediction to time series forecasting.
    
    \item \textbf{StockGAN} \cite{zhang2019stock} is a popular GAN model for stock prediction, which employs LSTM as the generator and Multi-Layer Perceptron (MLP) as the discriminator.
    
    \item \textbf{FactorVAE} \cite{duan2022factorvae} is the previous state-of-the-art model for stock market prediction, treating effective factors as the latent random variables in VAE.
    
\end{itemize}

Among baselines: Bollinger Bands is a momentum indicator in finance, BayesianNN, MQRNN, and CFRNN are
three most representative models for uncertainty quantification; StockGAN and FactorVAE are two popular generative models in the stock market which are able to produce sequences. Considering StockGAN and FactorVAE are point prediction methods and cannot be directly used for interval prediction, for a fair comparison, we develop StockGAN* and FactorVAE* which simply apply our proposed interval construction phase to their output sequences from the original models and build intervals.

\begin{table*}[!htp]
    \centering\fontsize{7.5}{7.5}\selectfont
    \caption{Numerical results regarding CP (\%), NMW(\%), and CWC(\%) for interval prediction, averaged over 5 independent runs. The improvement is significant statistically (t-test with $p<0.01$). }
    \begin{tabular}{c|ccc|ccc|ccc|ccc|ccc}
    \hline\multirow{2}*{Model} & 
    \multicolumn{3}{c|}{DJIA} & 
    \multicolumn{3}{c|}{SPX}& 
    \multicolumn{3}{c|}{Nasdaq}  &
    \multicolumn{3}{c|}{DAX} & 
    \multicolumn{3}{c}{Nikkei}\cr
    \cline{2-16}  &CP& NMW &CWC &CP& NMW&CWC &CP & NMW &CWC & CP & NMW & CWC & CP & NMW  & CWC\cr \hline
    Bollinger Bands & 81.97 & 7.52 & 21.95 & 81.97 &  6.07 & 17.71 & 81.56 & 5.13 & 15.18 & 83.32 & 10.74 & 30.00 & 82.90 & 10.13 & 28.68\\
    BayesianNN & 86.78 & 13.76 &34.51 & 88.64& 12.40 & 29.44 & 91.07 & 12.85 & 28.49 & 85.69 &  13.64 & 35.37  &86.46  & 12.68 & 32.11 \\
    MQRNN & 86.37 & 6.11 &15.52 & 85.60 & 5.39 & 14.01 & 95.00 & 6.63 & 6.63 & 84.89 & 8.86 & 23.57 & 86.88 &  5.03 & 12.58\\
    CFRNN & 99.86 & 56.01 & 56.01  & 100 & 62.51& 62.51& 100 & 25.70 & 25.70 & 100 & 23.09 &23.09 & 100  &  16.93  &  16.93 \\\hline
    StockGAN*   & 46.65 & 2.99& 36.53 &59.77 & 3.40 &23.19 & 60.24 &  3.68 & 24.60 &42.79 & 3.26 &47.61 & 36.83  & 3.39 &65.53\\
    FactorVAE* & 100 & 27.95 & 27.95 & 100 & 22.79 & 22.79  & 100&  17.35 &  17.35 & 100 & 36.52 & 36.52  & 100 & 32.01 & 32.01\\\hline
    % \RAGIC-risk &89.90& 4.31 & 9.87  & 91.52 &  3.92 & 8.59 & 92.81  & 3.30& 6.98 & 87.04 &  5.28 & 13.14 & 91.62 &  5.82 &12.71\\
    % \RAGIC-fixed & 92.67 & 4.80 & 10.19 & 90.91 & 3.37 & 7.48 & 89.90 & 2.82 & 6.46 & 91.90 &  6.60 & 14.30 & 93.86 & 6.85 & 14.10  \\
    \RAGIC & 95.56 & 5.42 & \textbf{5.42}  & 95.21 & 4.13 & \textbf{4.13}  &95.81 & 3.99 & \textbf{3.99}  & 95.81 & 7.78 & \textbf{7.78} & 96.28 & 7.96 & \textbf{7.96}  \\
    \hline
    \end{tabular}
    
    \label{tab:perf_nmw}
\end{table*}

\begin{figure*}[!htp]
\centering\includegraphics[width=.99\textwidth]{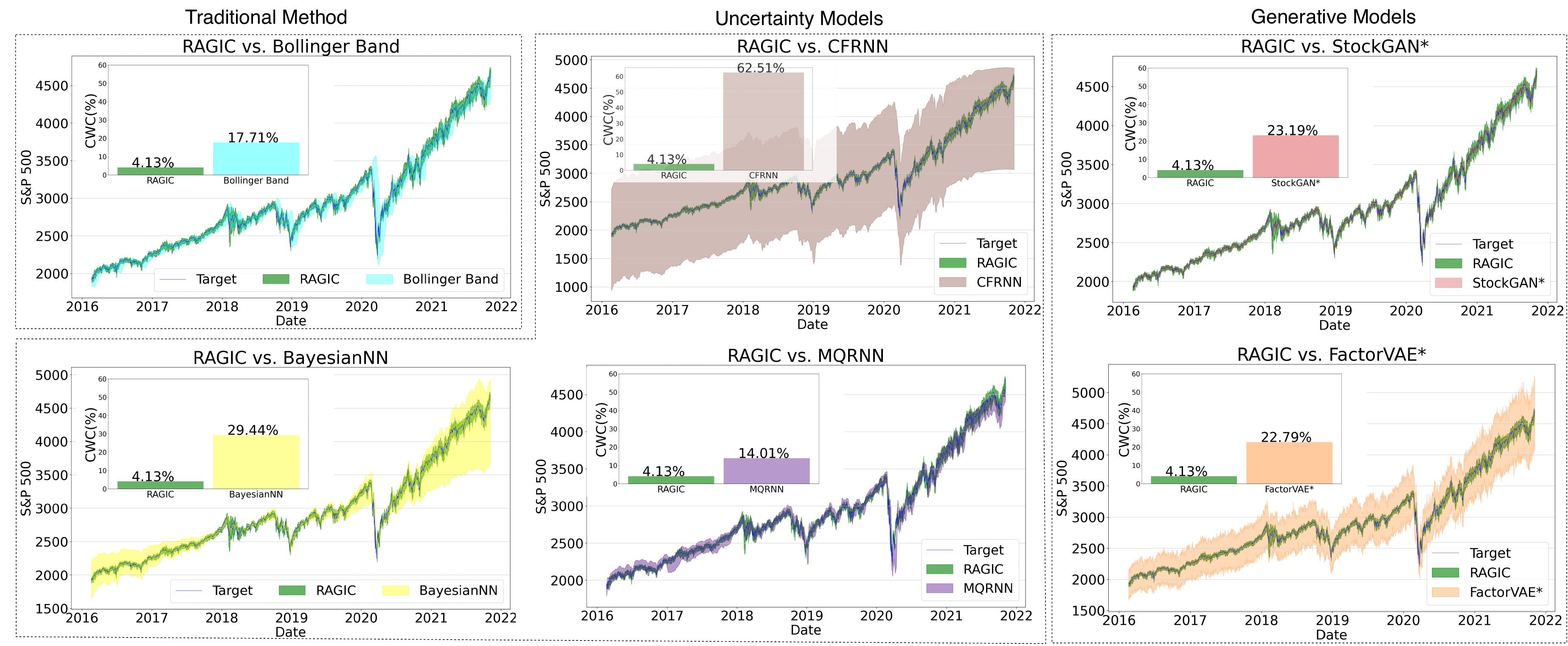}
  \caption{Comparison of predicted intervals on SPX.}
  \label{fig:interval}
\end{figure*}

\subsubsection{Evaluation Metrics}
We follow common practice to adopt three metrics to evaluate interval prediction performance:

\begin{itemize}
    \item \textbf{Coverage Probability (CP)} is the percentage of true values falling into the predicted interval over the test period.
    \begin{equation}
    CP = \frac{1}{T}\sum_{t=1}^T c_t, \quad 
    c_t =  
    \begin{cases}
    1 & y_t\in[Y_t^L, Y_t^U] \\
    0 & y_t\notin[Y_t^L, Y_t^U] \\
    \end{cases}
    \end{equation}
    
    \item \textbf{Normalized Mean Width (NMW)} is the mean width normalized by the range of close price in the test period, where mean width is the mean difference between upper and lower bounds.
    \begin{equation}
    NMW = \frac{1}{TR}\sum_{t=1}^T (Y_t^U- Y_t^L), \quad R = max(y)-min(y)
    \end{equation}

    \item \textbf{Coverage Width-based Criterion (CWC)} is a composite metric combining CP and NMW~\cite{khosravi2010lower}, showing the balance between accuracy and informativeness.
    
\begin{equation}
    CWC = NMW(1+\exp{(\eta\max(0, 95\%-CP))})
\end{equation}
$\eta$ is a penalty parameter on width if coverage does not achieve the expectation. A larger $\eta$ leads to an extremely higher CWC if coverage falls short of the desired 95\%, while a smaller value may not sufficiently highlight the difference. In our paper, we set it to 5 because it provides a more effective and straightforward model performance. A perfect interval should have high coverage, low width, and a low value of CWC. 

\end{itemize}

\subsubsection{Implementation Details:}
\RAGIC is implemented in PyTorch.  Hyperparameters are tuned based on the validation set. We use Adam as the optimizer with 50 epochs. The learning rate of critic is set to 0.0003. The threshold for weights clipping $\xi$ is set to 0.01. The number of generated sequences $N$ is 50. The penalty of supervised loss $\gamma$ is 0.3. The number of heads in multi-head attention is 2. The threshold $\Delta\epsilon$ in the Sigmoid curve is 0.0001. $\alpha_l$ and $\alpha_u$ are set to 90\% and 99.9\%. The batch size varies in \{150, 256\}. The learning rate of the generator varies in \{0.00005, 0.0001\}. In risk module, the risk threshold $\bm{\delta}_k$ for volatility return and normalized volatility index vary in \{0.05,0.1,0.2\} and \{0.05,0.1,0.2,0.25\} respectively, the coefficient $\bm{\lambda}_k$ vary in \{1.5,2.5\} and \{1.5, 2.5, 4\} respectively. In TCN, the number of layers $L$ varies in \{2,3\}, with the kernel size $p$ in \{5,8\} and the dimension of hidden layer in \{100,150\}. The frequency of critic update varies in \{3,4,5,6\}. The lower and upper threshlolds for volatility features in \emph{risk-sensitive interval} $v_l$, $v_u$ vary in \{10,12\} and \{20,22,25\} respectively.

For other baselines, we fine-tune the hyperparameters based on the optimal values reported in their works and choose the best model for implementation. Besides, below are two implementation differences between \RAGIC and baselines for fair comparisons:
\begin{itemize}
    \item Input features: Bollinger Bands uses close price only; other models are trained on the same features with \emph{RAGIC}
    \item Confidence $c$: the calculation of Bollinger Bands does not require it; BayesianNN, MQRNN, and CFRNN employ a target error rate which is equivalent to $1-c$ in our work, thus $c$ is set to 95\%; StockGAN* and FactorVAE* set the lower and upper bound of $c_t$ to 90\% and 99.9\%, same with \emph{RAGIC}. 
\end{itemize}

More details on the range of hyperparameters and the final chosen value of hyperparameters for \RAGIC and baselines on all stock indices are shown in the Supplementary.
Note that the optimal values shown in the appendix are provided as a reference for parameter settings that can lead to superior performance. \RAGIC has a high generalization ability so that multiple configurations can consistently produce good performance.

\subsubsection{Results and Discussion}
Table \ref{tab:perf_nmw} presents the numerical results, the hyperparameters are provided in Supplementary. We have the following observations. 
\begin{itemize}
    \item \RAGIC significantly outperforms other baselines in terms of CWC, which shows a balanced manner between accuracy and informativeness. It consistently achieves 95\% coverage on all indices with relatively narrow intervals.
    \item Intervals of \RAGIC, CFRNN, and FactorVAE* accurately contain true values with over 90\% of CP for all indices, while the width of \RAGIC is the narrowest (at least 5 times narrower than the other two for the U.S market). Hence, FactorVAE* and CFRNN are not competitive in terms of CWC. FactorVAE* maintains the perfect 100\% of coverage with over 15\% of NMW. We postulate the reason is that it cannot learn the posterior distribution well; hence individual point prediction is not robust and the range of interval with noise is large. CFRNN assumes exchangeable samples and adds the errors of the training set to the prediction of the test set, which does not fit the stock price time series data. For the task of stock market prediction, a narrow interval with high coverage is required, since an infinite interval that perfectly covers all possible prices is meaningless for decision-making.
    \item  StockGAN* produces the narrowest interval but fails with the lowest coverage. Possibly using LSTM as the generator can not well capture the complex temporal dependencies
    %, and one-step forecasting lacks possibility from multiple horizons. 
    %Grace: multiple step prediction is unique to us, so no need to mention others only use one-step forecasting
    \item For Bollinger Bands, BayesianNN, and MQRNN, the coverage achieves over 80\% while the width is larger than \RAGIC. Therefore, their performance regarding CWC is not ideal.

\end{itemize}

Moreover, Figure \ref{fig:interval} shows predicted intervals of benchmarks on S\&P 500 in the test period. The dark green area exhibits the interval of \RAGIC. It perfectly covers the actual price with a narrow interval, even in a very volatile market with a sharp drop, demonstrating the effectiveness of \RAGIC.

\subsection{Experiment 2: Ablation Study}

\subsubsection{Model components}
We implement three variants of \RAGIC to quantify improvement from the volatility index and temporal module.

\begin{figure}[!htp]
\centering\includegraphics[width=0.48\textwidth]{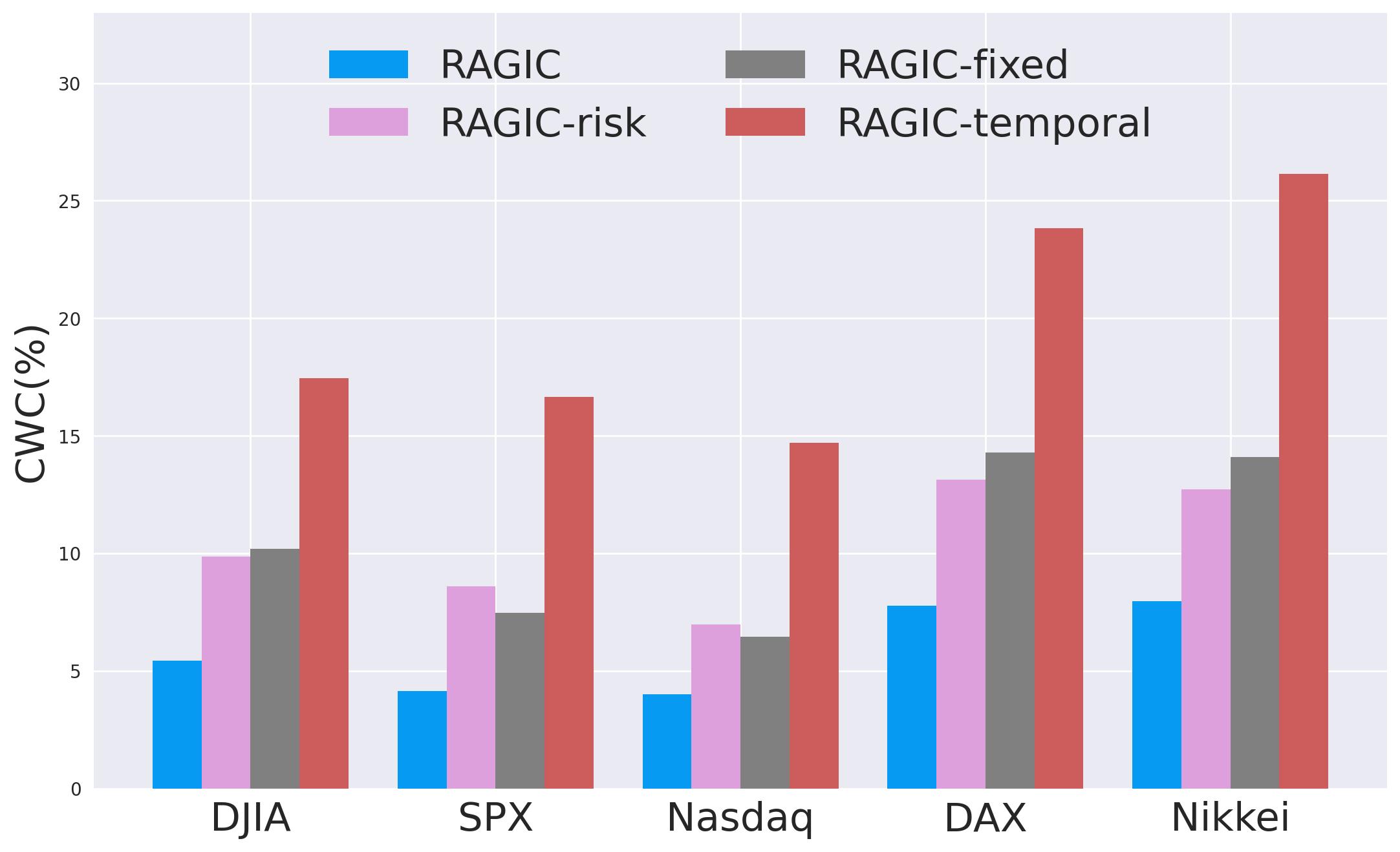}   
      \caption{Ablation Study.}
    \label{fig:ablation}
\end{figure}

\begin{itemize}
    \item \textbf{\RAGIC-risk} is a variant without the risk module to detect market risk and assign \emph{risk attention score} on stock features.
    \item \textbf{\RAGIC-fixed} is a variant fixing $c$ to 0.95 when building the interval.
    \item \textbf{\RAGIC-temporal} is a variant that replaces temporal module with a single fully connected layer.

\end{itemize}
The comparison of three variants on indices regarding CWC is visualized in Figure \ref{fig:ablation}. We observe that all three elements consistently contribute to the model quality on all indices. 
The volatility index in the risk module enhances performance by detecting high market risk; \emph{risk-sensitive interval} improves CWC by increasing coverage with
an acceptable sacrifice of width; the temporal module effectively mimics the dynamics of future patterns.

% \begin{figure}[!htp]
%     \centering      \subfigure{\includegraphics[width=0.23\textwidth]{images/ablation.jpeg} }  \subfigure{\includegraphics[width=0.215\textwidth]{images/horizon_cwc.png}}
%     \caption{Impact of model components (left) and prediction horizon (right).}
%     \label{fig:ablation}
% \end{figure}

\subsubsection{Prediction Horizon}
\begin{figure}[!htp]
    \centering    \includegraphics[width=0.48\textwidth]{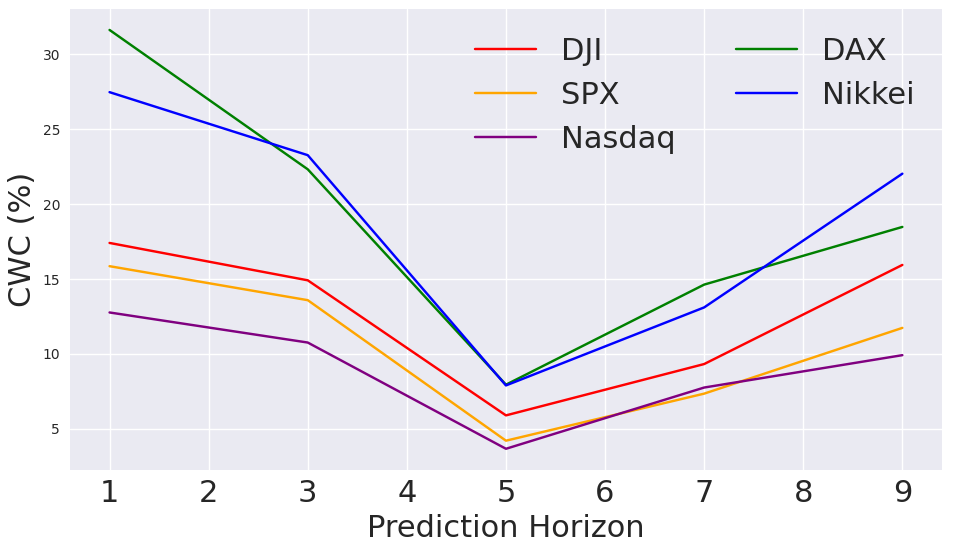}
    \caption{Impact of the prediction horizon.}
    \label{fig:horizon}
\end{figure}

We investigate the impact of prediction horizon $H$ on \RAGIC. 
Figure~\ref{fig:horizon} shows the change of CWC on different $H$.
As $H$ goes up from 2 to 5, the performance of the predicted interval shows a rising trend since richer information from a longer horizon is incorporated in building the interval, and the critic can differentiate real and fake sequences as their length becomes longer. CWC achieves the lowest value when $H$ is 5. 
However, as the length of the predicted sequence exceeds five days and continues increasing to 10 days, 
the interval quality goes down. It reveals that it becomes harder for GAN to simulate longer-term price patterns, and \emph{horizon-wise} information cannot provide more simulated uncertainty. 
The evaluation justifies 5 days forward as the optimal value for $H$. 
Also, note that 5 is the number of trading days in a week.

\subsubsection{Visualization of Risk Attention Score}
\begin{figure}[!htp]
    \centering   \includegraphics[width=0.5\textwidth]{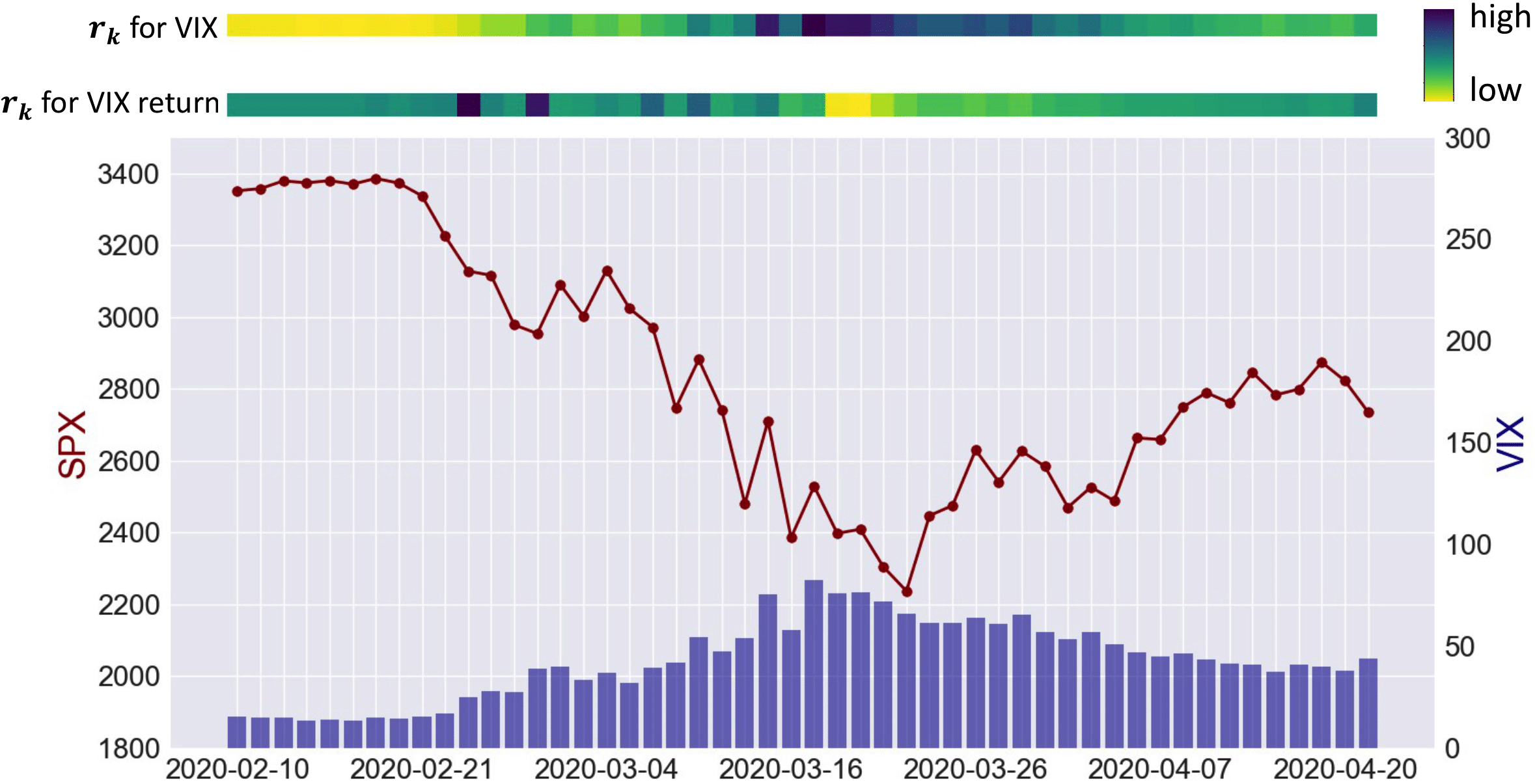} 
    \caption{Visualized risk score for VIX and VIX return} 
    \label{fig:risk_score}
\end{figure}

To verify that risk score can capture the market risk effectively, 
Figure~\ref{fig:risk_score} visualizes the risk attention score for two volatility-based features, VIX and the return of VIX, during the COVID pandemic. When the price begins to drop, the risk score captures the dramatic growth rate of VIX and alerts with an extremely high value on VIX return. During the crash period, the risk score for VIX remains high, which reveals that the investors lack confidence in the market. Both scores go down with the recovery of the market.

As $H$ goes up from 2 to 5, the performance of the predicted interval shows a rising trend since richer information from a longer horizon is incorporated in building the interval, and the critic can differentiate real and fake sequences as their length becomes longer. CWC achieves the lowest value when $H$ is 5.

\subsubsection{Various Methods of Adjusting Confidence $c$}
 
To quantify the improvement of employing volatility index on interval adjustment comprehensively, we implement  different values on various methods to determine $c_t$ for interval width.

\begin{table}[!htbp]
    \centering
    \caption{Comparison of interval width adjustment methods on DJI. Parameters for fixed value is $c$, 
    % parameters for cut-off value are $v_0$, $c_1$ and $c_2$, 
    parameters for sigmoid change are  $v_l$, $v_u$, $c_l$, and $c_u$.}
    \begin{tabular}{c|c|ccc}
    \hline
    Method & Parameters & CP & NMW  & CWC\\
    \hline
    \multirow{4}*{Fixed value} & 90  & 90.44 & 4.19 & 9.46 \\
    & 95 & 92.68 & 4.80 & 10.19 \\
    & 98 & 94.24 & 5.53 &  11.27 \\
    & 99.9 & 96.54 & 7.51  & 7.51\\
    \hline
    % \multirow{4}*{Cut-off value} & 10,90,99.9 &  96.54 & 7.44 & 7.44  \\
    % & 16,90,99.9  &  95.86 & 5.74 & 5.74 \\
    % & 22,90,99.9  &  93.76 & 4.79 &  9.88 \\
    %  \hline
    % Linear change & 10,25,90,99 & 94.92 & 1010 & 5.04 \\ \hline
    \multirow{2}*{Sigmoid change} & 10,22,90,99.9  & 95.66 & 5.42 & \textbf{5.42} \\
    & 10,22,95,99.9 & 95.86 &  5.82 & 5.82\\
    \hline
    \end{tabular}
    
    \label{tab:alpha}
\end{table}

Without loss of generosity, the experiment results on DJIA are shown in Table \ref{tab:alpha}. 
The value of $c_l$, $c_u$, $v_l$, and $v_u$ are set to 90\%, 99.9\%, 10, 22 respectively. 
We can see that simply fixing $c_t$ to a single higher value for all time steps in the testing period does not necessarily increase coverage, even though the interval gets wider. 
It is the least efficient method of increasing coverage and CWC in this table. 
% Cutting off $c_t$ at 16 of volatility produces the interval with slightly higher coverage but larger width compared with the sigmoid change method. 
It is worth noting that sigmoid change with a higher lower bound $c$ (95\%) covers more true close prices of DJIA over the test period, resulting in a higher coverage, while the interval gets slightly wider.
When this model is used, a user can make a decision between coverage and width given the application requirements.
%Therefore, we don't necessarily choose the model and parameters with the highest coverage. Such a trade-off between coverage and width is considered to a great extent. 

\subsection{Experiment 3: Point Prediction}

\subsubsection{Method} Besides generating an interval, \RAGIC can also compute the point prediction for price. Given the prediction vectors $\tilde{\bm{Y}_t}$ with element $\tilde{\bm{y}}_{t,n}=(\hat{y}^1_{t,n},..., \hat{y}^H_{t,n})$ from sequence simulation in the interval construction phase, we first compute weighted average value across the prediction horizon:
\begin{equation}
    \hat{y}_{t,n} = \frac{w_h\hat{y}^h_{t,n}}{\sum_{h=1}^H w_h}, \quad w_h=\frac{1}{2^h}
\end{equation}
where the weight $w_h$ exponentially shrinks as horizon step $h$ increases, considering the prediction from the most recent historical features is more reliable. Then we take the average over the repeated sequences as the final prediction $\hat{y}_t=\frac{1}{n}\sum_{n=1}^N \hat{y}_{t,n}$ for the price at time step $t$, which mitigates the noise caused by uncertainty in the financial market.

\subsubsection{Baselines} We identify ten baselines to compare with our work in point prediction. 
In addition to the eight models mentioned for interval prediction, we choose two recent popular works on stock point prediction. 
\begin{itemize}
    \item \textbf{HMG-TF} \cite{ding2020hierarchical} enhances Transformer with regularization and gaussian prior to predict stock.
    \item \textbf{AdaRNN} \cite{du2021adarnn} formulates the temporal covariate shift in non-stationary time series forecasting.
\end{itemize}

Among the previously mentioned eight models, StockGAN and FactorVAE are designed for point prediction and thus need not be adjusted in this comparison. 
The other six baselines for interval prediction are adjusted in the following way to be used for point prediction:

\begin{itemize}
    \item For Bollinger Bands, the middle point of upper and lower bounds is used as the prediction. 
    \item For BayesianNN, MQRNN, and CFRNN, we follow the original settings in their works \cite{stankeviciute2021conformal}. 
    \item For StockGAN* and FactorVAE*, the average of generated sequences for each time step is taken as the prediction. 
\end{itemize}

\begin{table}[!htbp]
    \centering 
    \caption{Numerical results regarding MAPE (\%) for point prediction, averaged over 5 independent runs. The improvement is significant statistically (t-test with $p<0.01$).}
    \begin{tabular}{c|c|c|c|c|c}
    \toprule 
    Model & DJIA & SPX & Nasdaq& DAX & Nikkei \\ \hline  
    Bollinger Bands & 1.92 & 1.90 & 2.43 & 2.16  & 2.24  \\
    BayesianNN & 1.72  & 1.86 & 3.29 & 1.65  & 1.29  \\
    MQRNN & 1.45 & 1.60  & 1.65  & 1.74 & 0.95 \\
    CFRNN & 2.21  & 2.25  & 1.74 &  1.87  & 1.85  \\\hline
    StockGAN* & 2.02  & 2.89 & 3.94  & 2.25  & 2.96 \\
    FactorVAE* & 1.01  & 1.03 & 1.13  & 1.00 & 1.07  \\ \hline
    HMG-TF & 0.76 & 0.71 & 0.91 &  0.84  & 0.93  \\
    AdaRNN & 0.79 & 0.87  & 0.93 &  0.83  & 0.94 \\ \hline
    StockGAN & 1.88 & 2.03 & 3.08 & 1.89 & 2.55 \\
    FactorVAE &  1.31  & 1.25  & 1.40  & 1.36  & 1.33 \\ \hline
    \RAGIC-risk & 0.85  & 0.85& 0.92 & 0.83  & 0.94 \\
    \RAGIC & \textbf{0.71}  & \textbf{0.69}  & \textbf{0.89} &  \textbf{0.81} & \textbf{0.89}\\\bottomrule
    \end{tabular}
    \label{tab: price}
\end{table}

\subsubsection{Results} We adopt Mean Absolute Percentage Error (MAPE) as the evaluation metrics and 
Table \ref{tab: price} shows the numerical results of the ten baselines and \RAGIC.
The table shows that \RAGIC outperforms all baselines for interval prediction regarding MAPE on all indices. Though FactorVAE* and CFRNN achieve full coverage, their point predictions are not accurate. The performance of FactorVAE on point prediction regarding MAPE is even worse, despite it's superior on ranking metrics\cite{duan2022factorvae}. It verifies our conjecture of large noise of individual point prediction which is mitigated in FactorVAE*. In addition, compared with baselines for point prediction on the stock market, \RAGIC is still competitive with the lowest MAPE.

\section{Discussion on Downstream Applications}
The interval prediction can offer investors valuable insights and aid them in making more informed and effective decisions. Its utility in downstream applications can be illustrated as follows:
\begin{itemize}
    \item Interval information can be heavily used in stock option trading. For example, if the predicted interval is [90,100], one profitable trading strategy is to sell a call option with a strike price $100+\epsilon$, and sell a put option with a strike price $90-\epsilon$. 
    \item The upper and lower bounds can serve as automatic take-profit and stop-loss levels for closing positions, which helps maximize the profit while mitigating risks effectively. 
    \item The width guides position sizing and risk management. A wide interval indicates higher risk, suggesting the need to reduce the position size to limit exposure, and vice versa. 
    \item When employing deep reinforcement learning for portfolio management, the prediction interval can be part of the state as a valuable reference for the model to determine the trading action. 
\end{itemize}

\section{Conclusions and Future Works}
This paper aims to tackle the challenge of stock prediction arising from the inherently stochastic nature of stock prices. To address this, we introduce a novel framework called \RAGIC. Our approach enhances the capabilities of the Generative Adversarial Network (GAN) in a unique manner, enabling the generation of a set of future price sequences with randomness derived from the financial market. Additionally, \RAGIC constructs a \emph{risk-sensitive interval}, which offers more comprehensive information compared to point predictions. Further experimental details can be found in Supplementary.

Looking ahead, our future endeavors will be focused on two key directions: (1) extending the generalization of \RAGIC to individual stock interval prediction, and (2) exploring downstream tasks to leverage the easily implemented \RAGIC for generating more profitable trading strategies to aid decision-making.

\bibliographystyle{IEEEtran}
\bibliography{reference}

\newpage
\begin{IEEEbiography}[{\includegraphics[width=1in,height=1.25in,clip,keepaspectratio]{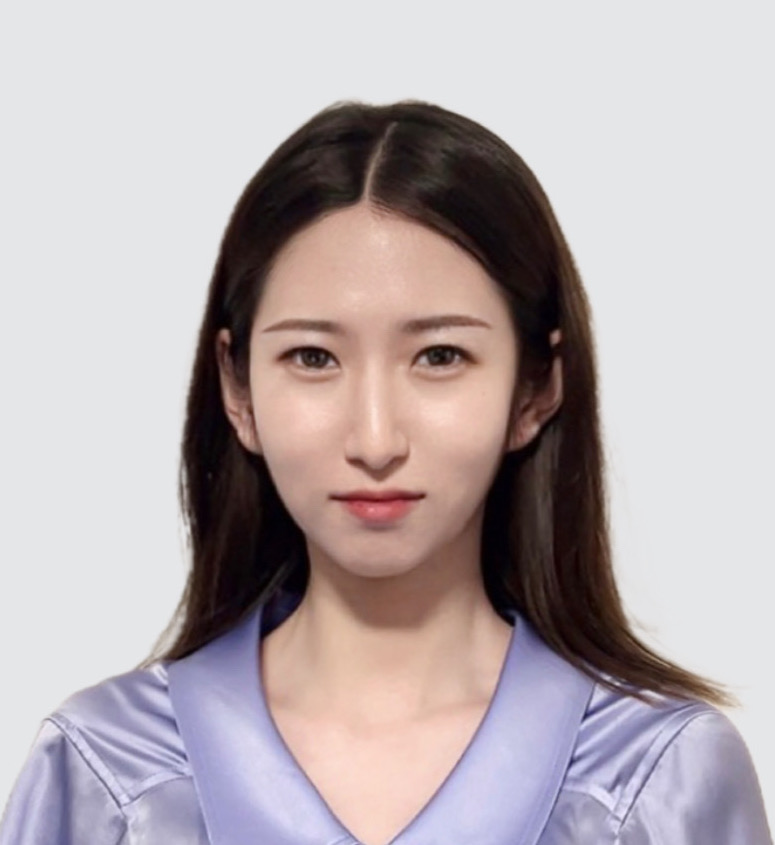}}]{Jingyi Gu} received the B.E. degree in Economics Statistics from Xiamen University, China, in 2017, the M.S. degree in Data Science from Rutgers University, New Brunswick, in 2019. She is currently pursuing the Ph.D. degree in Computer Science in New Jersey Institute of Technology. Her research interests include deep learning applications in various fields.
\end{IEEEbiography}

\begin{IEEEbiography}
[{\includegraphics[width=1in,height=1.25in,clip,keepaspectratio]
{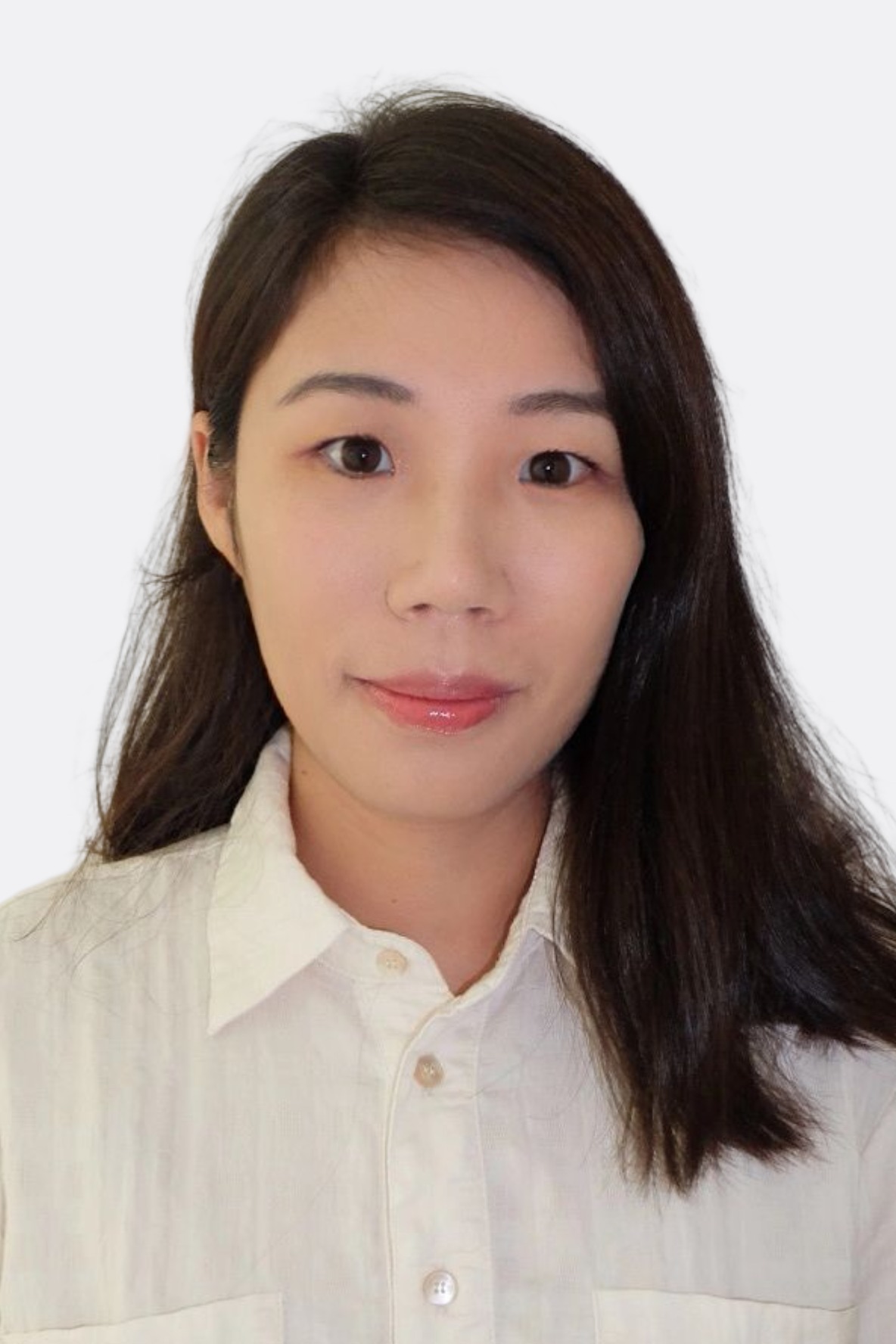}}]
{Wenlu Du} received the M.S. degree in Computer Science  from Worcester Polytechnic Institute, Worcester, in 2015. She worked as a software engineer in Alpine Electronics of America, Ohio, for two years. She is currently pursuing the Ph.D. degree in Computer Science in New Jersey Institute of Technology. Her research interests include reinforcement learning applications in various fields.

\end{IEEEbiography}

\begin{IEEEbiography}[{\includegraphics[width=1in,height=1.25in,clip,keepaspectratio]{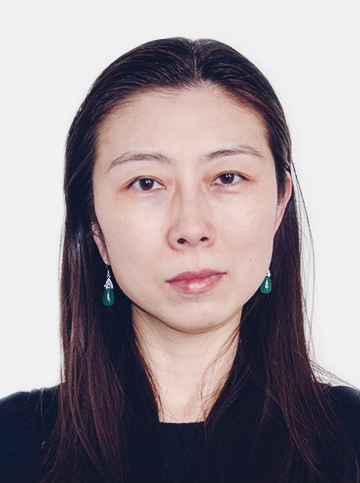}}]{Guiling (Grace) Wang}
(Fellow, IEEE) is currently a distinguished professor and the associate dean for research of Ying Wu College of Computing. She also holds a joint appointment at the Martin Tuchman School of Management and the Data Science Department. She received her Ph.D. in Computer Science and Engineering and a minor in Statistics from The Pennsylvania State University in 2006. Her research interests include FinTech, applied deep learning, blockchain technologies, and intelligent transportation.
\end{IEEEbiography}

\vfill

\end{document}